%% file: 00-main.tex
\pdfoutput=1
\documentclass[runningheads]{llncs}

\usepackage{comment}
\usepackage[hidelinks]{hyperref}

\usepackage{xurl}

\usepackage[T1]{fontenc}
\usepackage{graphicx}
\usepackage{color}

\begin{document}

\graphicspath{ {./figures/} }

\title{Overview of Web Application Performance Optimization Techniques}
\author{Juho Vepsäläinen\orcidID{0000-0003-0025-5540}, Arto Hellas\orcidID{0000-0001-6502-209X}, and Petri Vuorimaa\orcidID{0009-0007-6198-6650}}
\authorrunning{J. Vepsäläinen et al.}

\institute{Aalto University\\
\url{https://www.aalto.fi/en/department-of-computer-science}\\
\email{juho.vepsalainen@aalto.fi}}

\maketitle              % typeset the header of the contribution
%
% The abstract should briefly summarize the paper's contents in 150--250 words.
\begin{abstract}
During its thirty years of existence, the World Wide Web has helped to transform the world and create digital economies. Although it started as a global information exchange, it has become the most significant available application platform on top of its initial target. One of the side effects of this evolution was perhaps suboptimal ways to deliver content over the web, leading to wasted resources and business through lost conversions. Technically speaking, there are many ways to improve the performance of web applications. In this article, we examine the currently available options and the latest trends related to improving web application performance.

\keywords{Optimization \and Performance \and Web Applications \and Web Development \and Web Frameworks \and Web Tooling \and World Wide Web}
\end{abstract}

\vspace{10pt}

\textbf{The paper was accepted for Lecture Notes in Business Processing (LNBIP) and this version has been published with the permission of the publisher.}

\section{Introduction}
\label{sec:introduction}
\input{01-introduction}

\section{Aspects of Web Application Performance}
\label{sec:aspects}
\input{02-aspects}

\section{Review of Web Application Performance Techniques}
\label{sec:review}
\input{03-review}

\section{Impact of Web Frameworks on Performance}
\label{sec:frameworks}
\input{04-frameworks}

\section{Discussion}
\label{sec:discussion}
\input{05-discussion}

\section{Conclusion}
\label{sec:conclusion}
\input{06-conclusion}

% \pagebreak

\bibliographystyle{splncs04}
\bibliography{references}

\end{document}

%% file: 01-introduction.tex
The early web was meant as a global information exchange, and it came with critical concepts, such as hyperlinking, allowing documents to be linked to each other \cite{bernerslee1992}. Since then, the web has become the most extensive application platform as predicted by \cite{taivalsaari2011web}, reaching the majority of the population \cite{statistaInternetSocial}. Growth of the web platform has not been without its problems, as evidenced by the so-called dot-com bubble at the end of the 90s \cite{morris2008analysis}. In essence, the web has gone through multiple hype cycles \cite{dedehayir2016hype}. Depending on the viewpoint, the latest iteration of the web could be considered 3.0 or 4.0 \cite{europaPressCorner}. The crux of 4.0 is the seamless integration of physical and digital worlds. It has been estimated that the value of the new space might be up to €800 billion by 2030~\cite{europaPressCorner}.

Due to its importance for business, the web is a significant cornerstone of the global economy. However, the size of web pages keeps growing steadily, evidenced by a nearly 600 \% increase in the median size on mobile between 2012 and 2022~\cite{httparchivePageWeight}. The increase likely has to do with the growing complexity of the sites and user expectations. The growth has a concrete cost regarding wasted resources and lost conversions. To understand what can be done about the situation, we have phrased our research question for this article as follows: What are the currently available ways for improving web application performance?

To achieve this task, we first consider the aspects of web performance in Section \ref{sec:aspects} before reviewing currently available web performance techniques in Section \ref{sec:review}. Then we consider the impact of web frameworks on performance in Section \ref{sec:frameworks} and discuss our main observations in Section \ref{sec:discussion} before concluding the article with final remarks in Section \ref{sec:conclusion}.

% TODO: Explain the current main problems of web application performance (motivation) briefly and motivate the research question

% RQ: What are the current ways in which web application performance can be improved?

%% file: 02-aspects.tex
Traditionally, web services have been built using some variant of the client-server model \cite{oluwatosin2014client}. This arrangement relies on trust between the parties, and as an alternative, decentralized options, such as peer-to-peer computing, have been proposed \cite{korpal2023decentralization}. In this review, we will focus on the former as it is the current mainstream model used by the web as we consider the aspects of web application performance.

\subsection{What is performance}

When it comes to performance, it is good to consider what it consists of and what we mean by it. In the ideal case, clients can access a service they want without a noticeable delay. One could even go as far as to say that a good web experience should not differ from an application you might use on a desktop machine or a mobile phone. Web-based technologies such as \href{https://www.electronjs.org/}{Electron} are commonly used to develop desktop applications these days \cite{scoccia2020web}. \href{https://reactnative.dev/}{React Native} and \href{https://flutter.dev/}{Flutter} are comparable examples from the mobile side \cite{wu2018react}\footnote{To bridge the gap between native mobile experiences and the web, the concept of Progressive Web Apps (PWAs) was introduced by Google in 2015 \cite{tandel2018impact}. You could say a full parity can likely never be reached, but that is not to say there is no value in trying. PWAs enhance web applications through capabilities, such as push notifications, offline usage, and access via the home screen \cite{tandel2018impact}.}. Compared to these examples, the added requirement of the web has to do with the need for a server.

Web performance has been widely studied, as shown by \cite{aqeel2020landing}. \cite{killelea2002web} provides an excellent complementary snapshot of the topic and the techniques that were leveraged at the beginning of the millennium. As pointed out by \cite{aqeel2020landing}, most of the earlier research has focused on landing pages while ignoring internal pages, which is unfortunate as they can have differing performance. To give an example of a set of performance metrics leveraged by many developers, we will focus on Google's Core Web Vitals in this discussion as they capture the main aspects of web application performance.

According to \cite{webVitalsArticles}, Google evaluates the following six parameters known as Core Web Vitals to rank site speed: Largest Contentful Paint (LCP), First Contentful Paint (FCP), First Input Delay (FID), Interaction to Next Paint (INP), Cumulative Layout Shift (CLS), and Time to First Byte (TTFB). Suffice it to say that there are many ways to look at performance and to make things more complex; the ways to evaluate website speed change over time, as shown by the recent replacement of FID by INP in Google's Core Web Vitals \cite{webInteractionNext}. The reasoning for this change is that INP captures site responsiveness better than FID \cite{webInteractionNext}. Changes like this imply that web developers must know how their websites work against the currently relevant metrics to maintain their visibility in search results, as the metrics also contribute to search engine rankings, as discussed subsequently. Table \ref{table:web_vitals} explains the differences in the metrics in greater detail.

\begin{table}
    \centering
    \begin{tabular}{|p{4.5cm}|p{7.5cm}|}
        \hline
        Metric & Description \\
        \hline
        Largest Contentful Paint (LCP) & LCP measures loading performance, and for good UX, LCP must occur within 2.5 seconds of when the page starts loading \cite{webVitalsArticles}. \\
        \hline
        First Contentful Paint (FCP) & FCP measures the time it takes from when the user navigates to the page to when any part of the page's content is rendered on the screen \cite{webFirstContentful}. Time to Interactive (TTI) is a related metric that helps to identify cases where a page looks interactive but is not \cite{webTimeInteractive}. \\
        \hline
        First Input Delay (FID) & FID measures interactivity; ideally, it should be 100 milliseconds or less \cite{webVitalsArticles}. Note that INP has replaced FID and can be considered an obsolete \cite{webInteractionNext}. \\
        \hline
        Interaction to Next Paint (INP) & INP measures the overall responsiveness of a page by observing the latency of all click, tap, and keyboard interactions \cite{webInteractionInp}. \\
        \hline
        Cumulative Layout Shift (CLS) & CLS measures visual stability, and the value should remain 0.1 or less \cite{webVitalsArticles}. \\
        \hline
        Time to First Byte (TTFB) & TTFB measures the time between the request and when the first byte of a response begins to arrive \cite{webTimeFirst}. Occasionally, this is also known as Server Response Time (SRT). \\
        \hline
    \end{tabular}
    \caption{The table enumerates key differences between Google's Web Vital performance metrics \cite{webVitalsArticles}. It is worth noting that each metric looks at performance differently, and doing well in one does not mean you will do well in all of them.}
    \label{table:web_vitals}
\end{table}

\subsection{Search Engine Optimization}

Web Vitals are connected to Search Engine Optimization (SEO) as performance is a part of page ranking \cite{sellamuthu2022page}. Traditionally, SEO has meant optimizing websites for them to be easy for search engines to find, but as highlighted by \cite{eslamdoust2022overview}, the meaning has changed to search experience optimization, putting focus to the user instead of search engines as focusing on user has tangible benefits on the visibility of a website. Technically, SEO can be split into on-page SEO, off-page SEO, and technical SEO \cite{eslamdoust2022overview}. Technical SEO includes performance, and search engines evaluate it in various ways in their search ranking as a part of other factors. In other words, if you improve the performance of your website, you will likely improve its technical SEO as well.

\subsection{Transport protocol}

Starting from HTTP/2, the transport protocol of the web has been based on streaming, and the situation has evolved further with HTTP/3 as the protocol has moved to use UDP over TCP/IP underneath, allowing earlier limitations to be lifted \cite{vogel2023streaming}. As noted by \cite{vogel2023streaming}, even with a streaming protocol, performance is still limited by a phenomenon called render-blocking, which occurs when files that were streamed to the client have to be reassembled before the rendering of the page can continue; therefore, highlighting the need to avoid render-blocking operations at the client. As a potential solution, \cite{vogel2023streaming} proposes reorganizing page content and structure to be streaming-friendly. The benefit of this approach can be seen in a reduced FCP as the browser can get to the content faster \cite{vogel2023streaming}.

It can be argued that with the prevalence of streaming, earlier best practices, such as avoiding HTTP requests, may have to be re-evaluated \cite{souders2008high}. As shown by the example of \cite{vogel2023streaming}, gaining most of the benefit from streaming may require adaptation at the developer, or perhaps even framework, side to avoid the render-blocking problem.

% Another way to approach the problem of render-blocking is to defer loading of interactive content. A common technique is to apply \texttt{async} and \texttt{defer} attributes of script tags to override default blocking behavior \cite{netravali2016polaris}.

% * HTTP 1 vs. 2 vs. 3
% * streaming (Vogel etc.)
% * out-of-order streaming

\subsection{Offloading work}

Although JavaScript runtimes have been traditionally single-threaded, introducing new primitives, such as Web Workers, has enabled developers to overcome this limitation \cite{pan2015gray}. Current browsers support Web Workers well, and global support is around 98.4\% \cite{caniusequotwebWorkersquot}. As highlighted by \cite{pan2015gray}, client and server share computational tasks, and due to the ambiguity of what kind of logic to execute and where apart from security-related controls, \cite{pan2015gray} have dubbed this type of computation as gray computing. Gray computing comes with ethical challenges related to how much client resources should be consumed, and there are costs and benefits to be considered as gray computing techniques may enable benefits, for example, in website responsiveness \cite{pan2015gray}.

Service Worker (SW) is a specific type of Web Worker that can perform operations in the background, enabling, for example, offline support and push notifications \cite{chinprutthiwong2021service}\footnote{More specifically, Service Worker can act as a proxy between the traffic between the client and the server enabling techniques such as prefetching and even testing \cite{mswjsMockService}.}. SW is closely related to Progressive Web Applications (PWAs) as it is a vital part of allowing web applications that operate as their mobile counterparts \cite{tandel2018impact}. Given SW can intercept traffic heading to a website, it represents a new vector for attacks \cite{chinprutthiwong2021service}, highlighting that additions to the web platform may raise new concerns.

It is well understood that many websites depend on third-party resources, which comes with user tracking problems, for example, \cite{ikram2019chain}. Using third-party resources can also lead to render-blocking and increased resource usage \cite{builderIntroductionPartytown}. One way to solve the problem is to push third-party-related loading to Web Workers as illustrated by \href{https://partytown.builder.io/}{Partytown} \cite{builderIntroductionPartytown}. Partytown provides an excellent example of tooling that can help developers leverage the capabilities of modern web browsers with low effort.

\subsection{Impact of server latency on user experience}

With the server requirement of web applications, you get the latency problem, which has haunted the web since its early days \cite{viles1995availability}. Server latency is also visible in Google's Web Vitals \cite{webVitalsArticles} as a factor for SEO optimizers to consider. The negative impact of high latencies on User eXperience (UX) and conversion are well understood \cite{bai2017understanding,basalla2021latency}. To address the problem, Content Delivery Networks (CDNs) were introduced early on \cite{nygren2010akamai}. Since then, the approach has begotten the concept of edge computing \cite{satyanarayanan2017emergence} as a programmable generalization.

\subsection{Caching at different levels of development stack}

There are several ways how to overcome latency. The primary solution provided by edge computing is to push the server close to the user, as then there is less work to do by definition. That said, edge computing has additional needs, such as needing edge databases compatible with the globally distributed approach \cite{chen2021achieving}. Another approach to reducing latency is to decrease the amount of processing on the server side, for example, by caching \cite{song2012revisiting}. Latency can also be alleviated by sending smaller data payloads to the client and deferring the sending of data that is not needed immediately.

\begin{figure*}
    \centering
    \includegraphics[height=5.5cm, bb=0 0 2241 236]{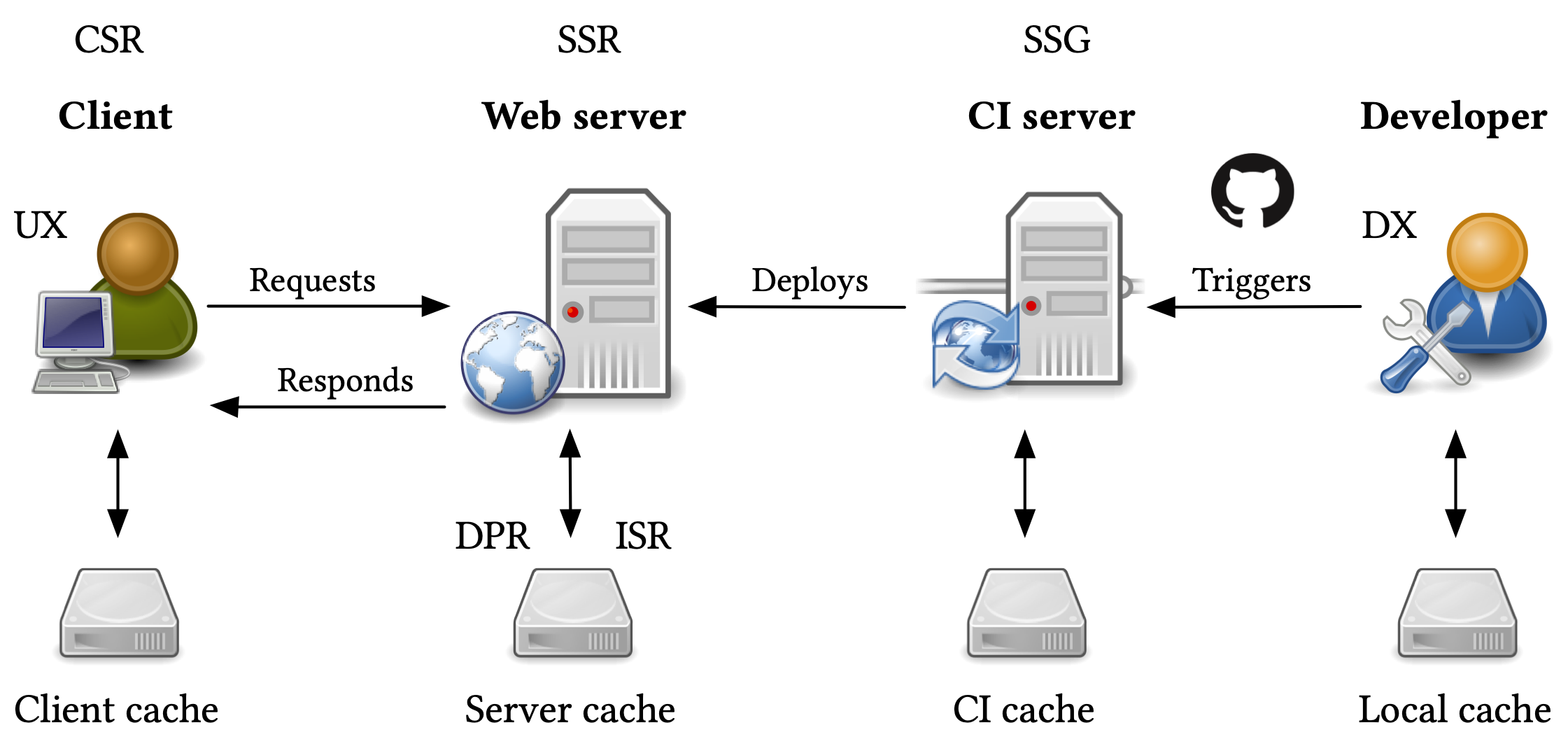}
    % \Description{The image shows a workflow from a client to a developer. The steps included are client, web server, CI server, and developer. Each step contains a cache. For the client, the concepts of UX and CSR are mentioned. For the web server, SSR and ISR are highlighted. At the CI server, SSG is mentioned as a related concept. For the developer, Developer eXperience (DX) is hinted at as a related concept.}
    \caption{In the workflow, a developer is working that triggers a build on a CI server, which deploys build artifacts to a web server that then serves a website to a client using a standard request/response flow. Caching is possible at each phase of the flow. In addition, a web server may generate a page per request using SSR or return it from a cache using DPR or ISR. During a CI deployment, SSG may be performed. Optionally, a page may be rendered on the client by leveraging CSR. \cite{vepsalainen2023implications}}
    \label{fig:client-dev-ssg}
\end{figure*}

Caching can be applied on different levels of a client-server arrangement, as seen in Figure \ref{fig:client-dev-ssg}. For example, clients can have a cache of their own, and there can be a server cache \cite{vepsalainen2023implications}. To complicate things further, the server cache can be populated differently. For example, in Distributed Persistent Rendering (DPR) or Incremental Static Regeneration (ISR), the server renders a snapshot of a requested page to the cache served from the cache for the subsequent requests \cite{vepsalainen2023implications}. DPR and ISR have slightly different takes on stale content as ISR can return stale results while DPR cannot \cite{vepsalainen2023implications}.

To further complicate things, code that runs on a server has to be deployed somehow. A static file server can be used for Static Site Generation (SSG), or a static website could be deployed on CDN infrastructure \cite{vepsalainen2023implications}. Server-Side Rendering (SSR) or Client-Side Rendering (CSR) could be leveraged for more complex cases \cite{vepsalainen2023implications}. In the case of SSR, the web server would generate requests on the fly and then optionally cache them as described above. Alternatively, a service such as Memcached or Varnish could be used between the client and the server \cite{song2012revisiting}.

Developers exist at the other extreme as they develop their code, most likely locally, and then push it to a version control system, which triggers a Continuous Integration (CI) server, which then deploys the built artifacts to a web server. The exciting implication of the flow is that a developer may experience the performance of a system differently locally than in a deployed environment due to differences in hardware and environment. To experience the impact of latency in the production environment, you would have to deploy it to equivalent infrastructure or simulate the conditions locally by, for example, throttling connection speed.

\subsection{Overview of aspects of web performance}

Given that the web platform has evolved substantially during the past thirty years, there are many opportunities for web developers and web framework authors to improve their work quality, especially regarding performance. Table \ref{table:aspects} recaps the directions discussed in the context of this article.

\begin{table}
    \centering
    \begin{tabular}{|p{2.0cm}|p{5.0cm}|p{5.0cm}|}
        \hline
        Aspect & Description & Related examples \\
        \hline
        Server latency & Server latency is a well-understood and studied problem, given that it cannot be avoided in traditional web applications. & CDNs \cite{nygren2010akamai} and edge computing \cite{satyanarayanan2017emergence} provide ways to address server latency by moving assets and computation closer to the client. With Local-First Software \cite{kleppmann2019local}, the issue can be alleviated by moving computation to the client and minimizing server roundtrips. \\
        \hline
        Transport protocol & As the web has evolved, so have the transport protocols underneath it. The biggest change has been the shift to streaming approaches \cite{vogel2023streaming}. & To leverage streaming fully, \cite{vogel2023streaming} experimented with reorganizing pages as streaming-friendly to avoid the problem of render-blocking. \\
        \hline
        Offloading work & Web platform allows performing secondary work within Web Workers \cite{pan2015gray} and Service Worker \cite{chinprutthiwong2021service} by moving it out of the main thread. & Partytown \cite{builderIntroductionPartytown} is a good example of a small library pushing third-party related work to Web Workers \cite{builderIntroductionPartytown}. \\
        \hline
        Caching & As a general technique, caching can be applied at different levels of a web application stack to reduce the work required. & On the server, Memcached and Varnish are well-known examples \cite{song2012revisiting}. On the client, the Service Worker can act as a caching proxy \cite{chinprutthiwong2021service}. \\
        \hline
    \end{tabular}
    \caption{The table discusses the main aspects of web application performance. As aspects, we highlight particularly server latency, transport protocol, offloading work, and caching as optimization directions.}
    \label{table:aspects}
\end{table}

% Table:
% Aspect | Optimization
% Code | Framework, library
% Delivery speed | Edge computing, local first
% Payload size | Compression, partial loading

% TODO: Include WebAssembly unless it is too specific - a way to optimize?

%% file: 03-review.tex
Building on the main aspects behind web application performance, we next focus on techniques for speeding up applications.

\subsection{Progressive enhancement}

Although it was not conceived as a performance-oriented technique but rather as a design to focus on the user, Progressive Enhancement (PE) provides a baseline for building site fundamentals first \cite{gustafson2015adaptive}. In PE, the idea is to start markup (i.e., HTML) first and use the correct semantics, build up styling (i.e., CSS), and only last develop interactivity (i.e., JavaScript) to follow the best practice of graceful degradation of the early web \cite{gustafson2015adaptive}. The crux of the approach is not to rely on the availability of CSS or JavaScript on the user side. New approaches, such as disappearing frameworks \cite{vepsalainen2023rise}, build on this idea and often provide a fallback that allows the usage of an application without JavaScript enabled. We argue that although the focus here is not on performance, it is more of a side effect of focusing on the fundamentals first, as the markup is the most valuable deliverable provided to the client while the rest follow.

\subsection{HTML-first development}

HTML First manifesto \cite{htmlfirstHTMLFirst} follows this line of thought as it argues that faster and more maintainable web applications can be built by leveraging the default capabilities of modern web browsers, the extreme simplicity of HTML's attribute syntax, and understanding there is power in the default behavior of web's ``view source'' functionality that allows developers to examine the markup of deployed applications. The manifesto aligns with the idea that modern web browsers are sufficiently advanced to be leveraged even without frameworks or perhaps with frameworks that do less, as shown by the examples of \href{https://enhance.dev/}{Enhance} \cite{enhanceEnhance} or \href{https://gustwind.js.org/}{Gustwind} \cite{gustwindGustwind}, frameworks that build on web standards while containing a few opinions of their own. The example of W3C's \href{https://open-ui.org/}{Open UI} \cite{openuiHomeOpen} Community Group shows developers' interest in improving the web platform so that more can be done directly without using third-party solutions for fundamental user interface operations.

\subsection{Utility-first styling}

Utility-first styling is a related approach in the sense that it is pushing style declarations to the HTML as styling tokens (e.g., \texttt{mx-2}\footnote{\texttt{mx-2} means that the browser should apply a margin of two units on the horizontal axis to the given element. Furthermore, the two units refer to a design scale that maps to the figure passed to the browser, offering a degree of abstraction and flexibility.}) that refer to associated CSS generated by a compiler \cite{klimm2021design}. Utility-first styling helps developers avoid traditional challenges of CSS by addressing issues such as specificity, resetting styles, location dependence, multiple inheritance, and nesting, to mention several \cite{klimm2021design}. Due to moving style declarations to the markup, it can be argued that utility-first styling may bloat HTML to some extent. Conversely, the generated CSS is generally a tiny amount compared to conventional techniques according to our experience with libraries like \href{https://tailwindcss.com/}{Tailwind} \cite{tailwindcssTailwindRapidly} or \href{https://twind.style/}{Twind} \cite{twindTwindstyle}.

\subsection{HTML-first state management}

In traditional web applications, so-called Multi-Page Applications (MPAs), the application state was maintained on the server side \cite{solovei2018difference}. Although the approach comes with its benefits in terms of simplicity and usable URLs\footnote{Uniform Resource Locators (URLs) give a way to identify unique resources available on the web through addresses \cite{akse2016applying}.} \cite{akse2016applying} for example, Single-Page Applications (SPAs) \cite{solovei2018difference} emerged to allow creation of complex web applications. Moving to the SPA model brought a new problem in the form of client-side state management \cite{akse2016applying}, and contemporary web application frameworks each address the issue in their own way.

Instead of solving state management through JavaScript, HTML-first state management approaches the problem from another angle by applying utility-first styling to the state. HTML-first state management solutions ask, what if the state was managed directly within HTML markup instead of separate JavaScript files? Generally, these types of solutions rely on a small runtime that makes state declarations associated with HTML alive, and they may also leverage techniques such as resumability \cite{vepsalainen2024resumability}. HTML-first state management is consistent with progressive enhancement, and due to its library-like nature, it can often be integrated within other frameworks, for example, to provide interactivity to an otherwise static page at a low cost.

\subsection{Local-first software – shifting responsibilities between the client and the server}

As pointed out by \cite{kleppmann2019local}, most current web applications are server-driven as a large part of the application state lives on the server. In local-first software, the client and server responsibilities are shifted, and the server gains a supporting role instead \cite{kleppmann2019local}. The key issues in distributed models like this are data ownership and the responsibility of processing. The interest in decentralization is visible in the advent of new social media platforms, such as Bluesky, that leverage new models of computation and state management \cite{kleppmann2024bluesky}.

\subsection{Partial loading}

Partial loading is a technique where content is loaded later \cite{netravali2016polaris} or, for example, on demand. As a more specific example, islands architecture allows developers to define interactive sections of a page and associate loading strategies to them \cite{vepsalainen2023state}. Through added control, developers can defer or even avoid loading functionality that may not be immediately needed. The concept of resumability~\cite{vepsalainen2024resumability} can be combined with a code-splitting~\cite{livshits2008doloto} approach to leverage the idea of partial loading on a granular level that was impossible earlier.

\subsection{Performance tooling}

While moving to a newer framework utilizing the latest techniques is an option, it may also be costly, especially for large codebases. Another orthogonal approach is to apply supporting tooling that makes the chosen framework faster. Examples of this direction are \href{https://million.dev/}{Million.js} \cite{bai2023million} and \href{https://github.com/waiter-and-autratac/WaiterAndAUTRATAC}{AUTRATAC/Waiter} \cite{vogel2023}. Million.js is an optimizing compiler for the popular virtual DOM, a technique utilized by libraries like React.js, and the compiler may lead to loading and rendering performance improvements up to orders of magnitude while providing a compatibility layer for React.js \cite{bai2023million}. In the approach pioneered by AUTRATAC and Waiter, the idea was to delay code execution by transpiling code to asynchronous code that can be downloaded later, resulting in reduced amount of render-blocking JavaScript as well as reduced amounts of downloaded code \cite{vogel2023}. The benefits of AUTRATAC/Waiter were visible in lowered FCP, especially in poor connection speeds \cite{vogel2023}, and in general, the same applies for techniques oriented at reducing initial payloads as there is less work to do then. \href{https://partytown.builder.io/}{Partytown} \cite{builderIntroductionPartytown} is another notable example as it allows web developers to push third-party dependencies, such as analytics, to be loaded through Web Workers in the background to avoid blocking JavaScript main thread.

\subsection{Dead Code Elimination}

% TODO: Connect DCE with the idea of minification
Another common approach to reducing the amount of JavaScript shipped to the client is Dead Code Elimination (DCE), and the target of this type of tooling is to analyze application code and drop unused parts from it \cite{malavolta2023javascript}. More specifically, these types of tools are called minifiers \cite{krol2020comparative}, and to maintain the debugging benefits of unminified code, typically, source maps are shipped in addition \cite{rack2023jack}. The risk is that minifiers can break a website by accidentally eliminating code that is needed, and that is a problem AUTRATAC/Waiter addresses by approaching the issue from a different angle \cite{vogel2023}. Tree-shaking is a DCE technique enabled by ES2015 modules, and the fact that it is possible to analyze which parts of them are used statically, but the limited adoption of the new standard may restrict the usefulness of this style of DCE \cite{obbink2018extensible}.

\subsection{Compression}

One way to reduce the amount of data sent to the client is to compress it before sending and decompress at the recipient \cite{ferragina2010compressing}. Many compression algorithms exist, where perhaps the most well-known options for web are gzip \cite{ferragina2010compressing} and Brotli \cite{alakuijala2015comparison}, which improves further compared to earlier gzip. According to \cite{caniusequotbrotliquotUse}, 97.52\% of users have access to Brotli through their browsers, meaning that applying Brotli compression is an easy way for developers to speed up their sites.

New image formats, such as WebP, promise better compression ratios over earlier formats, such as JPEG or PNG \cite{si2016research}. As with Brotli, the browser support for WebP is on a good level as 97.05\% of users have access to it \cite{caniusequotwebpquotUse}. The situation had significantly changed from 2015 when only roughly half of users had access to WebP, and its use cases were, therefore, more limited \cite{si2016research}. Due to the improved situation, WebP is a good option, especially for image-heavy applications. JPEG XL \cite{alakuijala2019jpeg} is another contender still being adopted by browsers, as shown by the early adoption rate of 10\% \cite{caniusequotjpegXlquot}.

% TODO: cost of parsing JavaScript from the resumability paper to a good spot

% Improved compression with visually good enough results is another promising direction especially for image heavy applications.

% * The idea is to send less to the client at the cost of having to do more work (decompression)
% * gzip/brotli
% * improved formats such as webp over jpeg/png (Jyrki Alakuijala)

% \cite{ferragina2010compressing} - gzip
% \cite{alakuijala2018brotli} - brotli
% \cite{alakuijala2015comparison} - brotli comparison

\subsection{Overview of performance optimizations}

Given that the web platform has evolved substantially during the past thirty years, there are many opportunities for web developers and web framework authors to improve their work quality by applying specific performance optimization techniques. Table \ref{table:performance} summarizes the directions discussed in this section.

\begin{table}
    \centering
    \begin{tabular}{|p{2.0cm}|p{5.0cm}|p{5.0cm}|}
        \hline
        Technique & Description & Related examples \\
        \hline
        Rendering approach & Web applications can be rendered either on the client or server side, and some of these techniques may complement each other \cite{vepsalainen2023implications}. & Most well-understood examples of rendering techniques are CSR, SSR, DPR, ISR, and SSG \cite{vepsalainen2023implications}. \\
        \hline
        HTML-first development & HTML-first development captures the sentiment related to using the web platform directly as much as possible. It builds on the earlier ideas of graceful degradation and progressive enhancement. & The best examples are HTML elements that can be used directly or techniques where CSS can be leveraged to provide interactivity without JavaScript. \\
        \hline
        Utility-first styling & Utility-first styling rethinks CSS by moving styling primitives to HTML markup. It may improve performance as a side effect due to smaller and more efficient CSS payloads. & Tailwind is perhaps the most well-known example of a utility-first library. \\
        \hline
        HTML-first state management & HTML-first state management solutions move state management directly to HTML markup and typically ship with a small JavaScript runtime, reducing the amount of code that must be shipped to the client. & \href{https://alpinejs.dev/}{Alpine.js} \cite{alpinejsAlpinejs} and \href{https://htmx.org/}{HTMX} \cite{htmxHome} are good examples with different takes on the topic, as Alpine.js is mainly frontend-focused while HTMX comes with backend opinions. \\
        \hline
        Local-first software & Local-first software re-evaluates how web applications should be built by shifting logic and data to the client and by treating the server in a supporting role \cite{kleppmann2019local}. & \href{https://github.com/automerge/trellis}{Trellis} \cite{githubGitHubAutomergetrellis}, a Trello clone, was an early example of the local-first approach and is still being explored as a technical option. \\
        \hline
        Partial loading & Partial loading allow developers to defer or even avoid work. & Islands architecture applies the idea and lets developers define islands of interactivity and associated loading strategies~\cite{vepsalainen2023state}. It is possible to leverage partial loading on top of resumability to a great degree~\cite{vepsalainen2024resumability}. Code-splitting~\cite{livshits2008doloto} is yet another example of how to apply partial loading on code. \\
        % Development techniques & Specific development techniques can be used to directly address aspects of web performance or as a side effect. & Good examples of web development techniques include code-splitting \cite{livshits2008doloto}, resumability \cite{vepsalainen2024resumability}, progressive enhancement \cite{gustafson2015adaptive}, HTML-first \cite{htmlfirstHTMLFirst}, utility-first styling \cite{klimm2021design}, HTML-first state management, Local-First Software \cite{kleppmann2019local}, and partial loading \cite{netravali2016polaris}. \\
        \hline
        Performance tooling & Performance-oriented tooling can provide benefits by integrating the tools into a project and then applying them. & Examples of performance-oriented tooling include AUTRATAC/Waiter \cite{vogel2023}, Million.js \cite{bai2023million}, and Partytown \cite{builderIntroductionPartytown}. \\
        \hline
        Dead Code Elimination & DCE can be used to eliminate code that the client does not need. Tree-shaking is a commonly used safe technique to achieve DCE. & Minifiers \cite{malavolta2023javascript} and tree-shaking bundlers \cite{obbink2018extensible}. \\
        \hline
        Compression & Given it is well understood that smaller payloads sent to the client result in better performance, compression techniques can be applied to the payloads. & Well-known compression techniques include \cite{ferragina2010compressing} and Brotli \cite{alakuijala2015comparison}. In addition, code can be minified to reduce its size \cite{krol2020comparative}, and new image formats, such as WebP \cite{si2016research} or JPEG XL \cite{alakuijala2019jpeg}, can be leveraged. \\
        \hline
    \end{tabular}
    \caption{The table discusses web application performance optimization directions for contemporary web developers. It is good to note that the list is likely non-exhaustive as new techniques become available constantly.}
    \label{table:performance}
\end{table}

%% file: 04-frameworks.tex
Developers often use a web framework for any application with a modest amount of complexity to save development time. The need for frameworks says something about the web platform itself. However, it can be argued that perhaps one day, frameworks won't be as needed as today as the platform covers enough functionality.

\subsection{Why web frameworks are needed}

Given that the web platform is powerful yet simultaneously complex, multiple web frameworks exist to capture the complexity and make it more approachable for developers. That is not to say there is no value in using the platform directly, and as the platform matures, there is likely less reason to use frameworks, at least for some use cases. Frameworks and their needs tend to inspire standards, and over time, the platform is catching up in terms of capabilities so that frameworks can do less or can do what they do even better.

\subsection{Server-, client-, and full-stack frameworks}

Generally, web frameworks can be split into server and client-oriented ones. Occasionally, a web framework covers both ends, but usually, there is a bias towards one end. So-called meta-frameworks go a step further by providing basic structure yet optionally requiring another framework to complement them. For example, \href{https://astro.build/}{Astro framework} \cite{astroAstro} has grown popular partially thanks to its meta-approach as it has enabled developers to bring familiar tools to a new environment with its unique benefits. \href{https://www.phoenixframework.org/}{Phoenix framework} \cite{phoenixframeworkPhoenixFramework} built using \href{https://elixir-lang.org/}{Elixir} \cite{elixirlangElixirProgramming} is a good example of a server-oriented web framework. \href{https://www.djangoproject.com/}{Django} \cite{djangoprojectDjango} developed with \href{https://www.python.org/}{Python} and \href{https://rubyonrails.org/}{Ruby on Rails} \cite{rubyonrailsRubyRails} are two more examples of a framework starting server-first. Astro framework has a heavier focus on the front end, but technically, it is possible to build full-stack applications using it. \href{https://nextjs.org/}{Next.js} \cite{nextjsNextjsVercel}, a full-stack framework built around Meta's \href{https://react.dev/}{React} \cite{reactReact}, is another example that can cover both ends.

\href{https://svelte.dev/}{Svelte} \cite{svelteSveltex2022}, \href{https://vuejs.org/}{Vue} \cite{vuejsVuejs}, and \href{https://angular.io/}{Angular} \cite{angularAngular} are examples of client-oriented frameworks. However, full-stack solutions are available for each, allowing the development of complex applications around them. As mentioned earlier, solutions like Astro allow the integration of these frameworks within another meta-framework.

The mentioned frameworks are only a few out of many, and you might ask why so many exist. Generally, frameworks capture several opinions on everyday tasks, such as routing, and exist to make these tasks more accessible for developers to perform. Occasionally, frameworks can help to pinpoint issues in platforms. For example, today's web frameworks can leverage standard features inspired by earlier frameworks and perhaps do less work independently.

\subsection{Impact of web frameworks on performance}

Given frameworks capture opinions, they can also help establish a baseline for web performance. Historically, SPA frameworks have focused on Developer eXperience (DX), perhaps at the cost of UX \cite{vepsalainen2023rise}. With the advent of disappearing frameworks \cite{vepsalainen2023rise}, it seems change is in the air, and techniques, such as resumability~\cite{vepsalainen2024resumability}, enable developers to write code that follows the currently understood best practices and is performant out of the box without the need for further manual optimizations, such as code-splitting where code is separated into multiple chunks that can be loaded for example based on user interaction \cite{livshits2008doloto}.

Conversely, a framework can work against developers by leading to technical costs that are not always apparent. For example, hydration, a common technique to turn pages interactive, comes with a loading cost that may not always be apparent to developers \cite{vepsalainen2024resumability}. Therefore, adopting a framework does not automatically mean an application will be more performant; the opposite may be true.

\subsection{Adoption of performance optimizations in web application frameworks}

To give a fuller view of the topic, Table \ref{table:performance-adoption} shows how web application frameworks have adopted performance optimization directions discussed in this article. Refer to Table \ref{table:performance} for brief explanations of the techniques.

\begin{table}
    \centering
    \begin{tabular}{|p{3.3cm}|p{8.7cm}|}
        \hline
        Technique & Relationship with web application frameworks \\
        \hline
        Rendering approach & Each web application framework is forced to implement some rendering approach by definition. Pure client-side frameworks implement CSR, while full-stack and server-side frameworks can leverage SSR, DPR, ISR, and SSG. \\
        \hline
        HTML-first development & HTML-first development approach has been adopted specifically by Enhance \cite{enhanceEnhance} and Gustwind \cite{gustwindGustwind}. \\
        \hline
        Utility-first styling & Utility-first styling is a technique that may be used by a framework through external libraries and tooling. Typically styling approach a framework is up to the developer to decide. \\
        \hline
        HTML-first state management & HTML-first state management are similar to styling in that, so far, frameworks have not adopted the idea. Instead, it can be implemented on top of them. A notable exception of deeper integration is Laravel Livewire, which includes first-class support for Alpine.js \cite{laravelAlpineLaravel}. \\
        \hline
        Local-first software & So far, no known example of a local-first framework exists as the approach is meant to be used on top. That is not to say there might not be value in a framework focusing solely on the approach as Astro did with islands architecture~\cite{vepsalainen2023state} earlier. \\
        \hline
        Partial loading & Partial loading has been adopted by frameworks, such as Astro~\cite{vepsalainen2023state}, Qwik through resumability~\cite{vepsalainen2024resumability}, and numerous other frameworks where it is possible to use code-splitting. Notably, in the latter case, significant developer effort may be required to gain major benefits. \\
        \hline
        Performance tooling & Performance tooling exist separately from web application frameworks; typically, it is not shipped with them. \\
        \hline
        Dead Code Elimination & DCE is commonly used in frameworks that leverage a bundler in their build process. A good example is the integration of minification to Next.js in production mode \cite{githubMinifyNext}. \\
        \hline
        Compression & Compression is typically exposed to the developer for full-stack or server frameworks. For example, in Next.js, compression is available through configuration \cite{nextjsNextconfigjsOptions}. \\
        \hline
    \end{tabular}
    \caption{The table combines web performance optimization techniques discussed in this article with web application frameworks to show their relationship. Notably, techniques such as applying different rendering techniques, partial loading, or applying performance tooling, DCE, and compression are common among frameworks.}
    \label{table:performance-adoption}
\end{table}

%% file: 05-discussion.tex
Optimizing web application performance is not an easy task as there are many different ways to approach the problem. Most likely, some Pareto principle \cite{dunford2014pareto} applies for many cases where a considerable performance improvement may be gained through a small amount of effort. Given the size of websites these days, addressing the size problem \cite{httparchivePageWeight} may be an excellent place to start.

\subsection{Addressing size of websites}

Some aspects discussed in this article, particularly web frameworks, development techniques, tooling, and compression, touch on website size. The interesting question is, what is the contribution of each factor to the whole, and how much can be done about it by improving a specific section or by, for example, applying a particular technique?

Perhaps the bigger underlying question is why websites are so large in the first place and what the main reasons behind their growth are. Even with enhanced approaches, could the growth be at least slowed down, or is there something more fundamental going on driving the development?

\subsection{Towards lower latency and distributed models}

A central underlying theme we noticed was the shift in the server space. The problem of server latency is well understood, so edge computing has gained interest in the web space during the past few years as commercial offerings keep maturing. Simultaneously, there has been interest in offloading work to the client and developing distributed models for web applications, as shown by local-first software \cite{kleppmann2019local}. The more prominent underlying theme in this rethinking has to do with aspects of privacy and data ownership.

\subsection{Evolutionary pressure of web frameworks}

To develop complex web applications, developers often resort to web frameworks that capture opinions, which avoids effort and hopefully enables best practices. The problem is that best practices tend to change over time, and the question is whether mainstream frameworks keep up with the change. Another evolutionary pressure comes from the standardization side as standards keep capturing critical ideas to the web platform itself, meaning there is less work for frameworks to do. A good example is the introduction of the ES2015 version of JavaScript language, which is now available in web browsers without external tooling. Also, the introduction of standards like Web Workers has opened new venues for technical development and rethinking best practices as new solutions, and even paradigms, can be developed on top of them.

\subsection{The way performance is measured shapes performance}

Undoubtedly, what you measure is what you get, particularly when the measurements are used as metrics for decisions such as page rankings. Therefore, it has been interesting to notice that, for example, Google Web Vitals still keeps evolving, as shown by the introduction of INP \cite{webInteractionInp} as a web vital in favor of FID. Since web frameworks follow performance metrics closely to remain competitive, improving how we measure performance is likely good for the web platform overall.

%% file: 06-conclusion.tex
In this article, we sought to find answers to the question \textit{What are the currently available ways for improving web application performance?} and found surprisingly many ways. At the core of it all is the aspect of measuring web performance, and it turns out Google's Core Web Vitals provide a good approximation of that. With the advent of edge computing, some focus has shifted to minimizing server latency. With the introduction of local-first software, there is interest in distributed models that move computation and data to the client, leaving servers more to a supporting role.

Web frameworks allow codifying best practices, but the concern is that although they save developer effort this way, they might also remain on the conservative side, at least when there are many developers to support, meaning it is up to the bleeding edge frameworks to innovate and try new techniques to gain market share. A more specific study into the way ideas between mainstream and bleeding edge frameworks transfer would be worthwhile as that would give more insight within the space to understand how frameworks adopt new standards and techniques that become available. Another direction of inquiry would be to consider how standardization can capture innovation happening at frameworks and how this affects the whole ecosystem.

In the scope of this article, we did not consider which factors make the most sense for individual developers to apply in their daily work. There are also questions related to which techniques web framework authors should adopt to disseminate current best practices to their userbases. It is a good question about which practices should be encoded into the frameworks and which should be applied by the users on top of them. Another related issue has to do with the web platform and its adoption of worthwhile techniques to raise the baseline for frameworks.